\documentclass[aps,prc,superscriptaddress,amsmath,amssymb,twocolumn, longbibliography]{revtex4-2}

\usepackage[utf8]{inputenc} 
\usepackage{graphicx}      
\usepackage{amsmath}       
\usepackage{amssymb}       
\usepackage{hyperref}      
\usepackage{natbib}        

\hypersetup{
    colorlinks=true,
    linkcolor=blue,
    citecolor=blue,
    urlcolor=blue
}

\begin{document}

\title{Comparative Study of Langevin and Random Walk Models\\
for Nuclear Fission in the Overdamped Regime}

\author{A. Augustyn}
\affiliation{National Centre for Nuclear Research, Pasteura 7, 02-093 Warsaw, Poland}

\author{T. Cap}
\affiliation{National Centre for Nuclear Research, Pasteura 7, 02-093 Warsaw, Poland}

\author{M. Kowal}
\email[Corresponding author: ]{michal.kowal@ncbj.gov.pl}
\affiliation{National Centre for Nuclear Research, Pasteura 7, 02-093 Warsaw, Poland}

\author{K. Pomorski}
\affiliation{National Centre for Nuclear Research, Pasteura 7, 02-093 Warsaw, Poland}


\date{\today}

\begin{abstract}
We present a comparative study of Langevin dynamics and a 
Metropolis random walk model applied to thermal 
neutron-induced fission of $^{229}$Th, $^{235}$U, 
$^{239}$Pu, $^{245}$Cm, $^{249}$Cf, and $^{255}$Fm. Both methods are 
implemented within an identical four-dimensional 
Fourier-over-Spheroid framework, using potential 
energy surfaces derived from the macroscopic-microscopic 
model. We show that the Metropolis walk corresponds to 
the overdamped limit of the Langevin equations and 
confirm this correspondence numerically by Langevin 
calculations performed in the strongly damped regime and 
with quantum corrections to the random force switched 
off. Under these conditions, the two approaches produce 
essentially identical mass distributions for the lighter 
actinides. Systematic deviations develop for the heavier actinides, where the 
Langevin dynamics yields a non-negligible symmetric fission component 
absent in the random walk results. We trace this 
difference to the kinematic structure of the Metropolis 
sampling and to the residual inertial dynamics retained 
in the Langevin framework. A parallel comparison of Langevin calculations with and without the 
quantum-corrected effective temperature $T^*$ isolates the contribution 
of zero-point fluctuations and suggests that their standard 
phenomenological treatment may overestimate their impact in certain cases.
Both approaches qualitatively reproduce the 
asymmetric peak positions and their systematic evolution 
across the actinide chain, while a common quantitative 
limitation --- the narrowness of the predicted 
distributions --- points to the role of 
higher-dimensional deformation modes not included in the 
present parametrization.
\end{abstract}
\maketitle 


\section{Introduction}
\label{sec:intro}
Nuclear fission, the process by which a heavy nucleus splits into lighter 
fragments, is a cornerstone of nuclear physics, pivotal for the fundamental 
understanding of nuclear structure, forces, and many-body dynamics, as well 
as for technological applications. This transformation involves intricate, 
large-scale collective motion, where the nucleus evolves through a sequence 
of complex shapes defined within a multi-dimensional deformation space. 
The dynamics are inherently dissipative, signifying a continuous energy 
exchange between the macroscopic collective coordinates defining the shape 
and the microscopic intrinsic degrees of freedom of the constituent 
nucleons~\cite{Krappe2012, Talou2023, NerloPomorska2024book}. 
Despite decades of research and significant advancements 
in theoretical modeling---ranging from microscopic theories based on energy 
density functionals~\cite{pei2009, schunck2014, bulgac2016, tao2017, zhao2020prc} to well-established macroscopic-microscopic 
approaches coupled with dynamical frameworks~\cite{asano2004, aritomo2014, sierk2017, ishizuka2017, usang2017, liu2019prc, 
pomorski2020prc, pomorski2021, pomorski2024}
---a complete microscopic description and full quantitative 
prediction of all fission observables remain challenging. 
Key open questions persist regarding the precise evolution pathways, especially 
during the descent from the saddle region to scission, 
as well as the detailed nuclear shape and the nascent fragment 
mass and charge partitions at the moment of rupture.

Microscopic approaches, such as time-dependent density functional theory 
(TDDFT), offer a path towards a first-principles description, providing 
insights into non-equilibrium dynamics, fission timescales, and fragment 
properties without adjustable parameters~\cite{pei2009, schunck2014, bulgac2016, tao2017, zhao2020prc}. However, these 
approaches face significant computational demands and encounter difficulties 
in consistently incorporating thermal fluctuations and dissipation at finite 
excitation energies.

To overcome these limitations, stochastic methods present a powerful 
alternative by explicitly modeling the interplay between collective motion 
and the intrinsic heat bath. Models like Langevin dynamics~\cite{asano2004, aritomo2014, sierk2017, ishizuka2017, usang2017, liu2019prc, 
pomorski2020prc, pomorski2021, pomorski2024} or Brownian 
shape motion~\cite{randrup2011, Randrup2011_PRC84, Randrup2013, MollerRandrup2015_Brownian, Albertsson2019EPJConf, Albertsson2020EPJA, mumpower2020}, 
simulated on detailed potential energy surfaces, have 
proven highly effective. 
Multi-dimensional Langevin dynamics simulations, incorporating inertia, potential forces, friction, and random forces, successfully describe the diffusive evolution across complex energy landscapes, while simpler random walk or Brownian motion models, focusing on phase-space exploration governed by level densities and potential barriers, help understand fission fragment mass yields and the competition between fission modes. The random walk approach has also been extended to the superheavy region to calculate fragment mass yields and total kinetic energies~\cite{Albertsson2020EPJA}.

The relevance of dissipative dynamics extends beyond fission itself. 
Analogous stochastic descriptions based on the Langevin and 
Fokker-Planck equations have proven indispensable 
for interpreting experimental observables in deep-inelastic collisions 
and fusion-fission processes, including cross sections, the degree of 
energy dissipation, and the mechanisms governing mass and charge transfer 
between interacting nuclei~\cite{Feldmeier1987, Froebrich1998}. These approaches also provide valuable insights into 
quasifission and the evolution of the nuclear system during the fusion 
process.

The aim of this work is to present a direct comparison between two 
distinct stochastic approaches to induced fission: multi-dimensional 
Langevin dynamics and the random walk methodology, which can be regarded 
as a limiting form of the Langevin equations in the overdamped regime 
(i.e., in the limit of large friction). 
Building on recent results obtained with a Langevin model using the 
Fourier-over-Spheroid shape parametrization~\cite{pomorski2023, pomorski2024}, 
both approaches are applied on identical 4D 
FoS-parameterized potential energy surfaces 
computed for a set of neutron-induced fission reactions on selected 
actinide targets: $^{229}$Th, $^{235}$U, $^{239}$Pu, $^{245}$Cm, $^{249}$Cf,
and $^{255}$Fm. This ensures full consistency of the underlying nuclear 
structure input across all studied systems. The unified computational 
setup allows for a rigorous evaluation of the similarities, differences, 
and underlying physics captured by each method.

\section{Common Theoretical Framework: Fourier-over-Spheroid Shape Parametrization and Potential Energy Surfaces}
\label{sec:framework}

Within the macroscopic-microscopic approach, deformation parameters serve 
as coefficients in the nuclear shape expansion, coherently linking the 
macroscopic liquid-drop energy with microscopic shell and pairing 
corrections to form the potential energy surface. Since no universal 
shape parametrization can optimally describe both compact mononuclear and 
strongly deformed binary nuclear configurations, the choice of 
parametrization must be suited to the deformation regime of interest. 
In this work, we employ the Fourier-over-Spheroid (FoS) parametrization 
owing to its mathematical flexibility and computational efficiency in 
modeling the elongated and mass-asymmetric shapes typical of nuclear 
fission~\cite{pomorski2023, pomorski2024}. The four collective deformation 
parameters used in this work are $c$, $a_3$, $a_4$, and $\eta$, where $c$ 
describes the relative elongation of the nucleus, $a_3$ and $a_4$ are 
the leading mass-asymmetry and necking parameters, respectively, and 
$\eta$ represents the degree of triaxial (non-axial) deformation.

In the FoS parametrization, the nuclear surface in cylindrical coordinates 
$(\rho, \varphi, z)$ is defined by~\cite{pomorski2023}:
\begin{equation}
\rho_s^2(z, \varphi) = \frac{R_0^2}{c} f\left(\frac{z - z_{\text{sh}}}{z_0}\right) 
\frac{1 - \eta^2}{1 + \eta^2 + 2 \eta \cos(2 \varphi)},
\label{eq:fos_rho}
\end{equation}
where $R_0$ is the radius of the spherical shape, $z_0 = c R_0$, and 
$z_{\text{sh}} = \frac{3}{2\pi} z_0 (a_3 - a_5/2 + \cdots)$ is a shift 
ensuring that the center of mass remains at the origin. The axial shape 
function $f(u)$ is given by:
\begin{equation}
f(u) = 1 - u^2 - \sum_{k=1}^n \left\{ a_{2k} \cos\left(\frac{2k-1}{2} 
\pi u\right) + a_{2k+1} \sin(k \pi u) \right\},
\label{eq:fos_f}
\end{equation}
with $u = (z - z_{\mathrm{sh}})/z_0$, where the coefficient 
$a_2 = a_4/3 - a_6/5 + \cdots$ is determined by the volume 
conservation condition. The parameters $a_5$, $a_6$, and 
higher-order terms account for additional multipole 
deformations. The effect of these deformations on static 
properties such as fission barriers and ground-state masses 
was discussed in~\cite{AugustynPRC}. The deformation 
parameter $a_6$ can play a significant role in actinide 
nuclei with particularly narrow ground-state minima, 
affecting fission barrier heights, although its influence 
on saddle-point energies is comparatively smaller. The 
parameters $a_5$ and $a_6$ modify the shapes of the 
fission fragments but not the overall elongation of the 
nucleus, and can therefore influence the detailed 
properties of the scission configurations, including the 
final mass split. In the vicinity of the scission point, 
however, restricting the parametrization to the three 
collective coordinates $c$, $a_3$, and $a_4$ has been shown 
to reproduce the gross features of fission fragment mass 
distributions and their kinetic 
energies~\cite{pomorski2024,pomorski2023}. The 
four-dimensional FoS parametrization adopted here therefore 
provides a sufficient basis for the present dynamical 
study, while a full exploration of the role of higher 
multipoles is deferred to future work.

The potential energy surfaces (PES) of fissioning nuclei are evaluated 
in the four-dimensional deformation space spanned by the collective 
coordinates $\mathbf{q} = \{c, a_3, a_4, \eta\}$ using the 
macroscopic-microscopic (macro-micro) 
model~\cite{Myers1966,Nilsson1969}:
\begin{equation}
V(\mathbf{q}, T) = V_{\text{mac}}(\mathbf{q}) + V_{\text{mic}}(\mathbf{q}, T),
\label{eq:pes_total}
\end{equation}
where $V_{\text{mac}}(\mathbf{q})$ is the 
macroscopic part of the energy, evaluated according to the 
Lublin-Strasbourg Drop (LSD) formula~\cite{Pomorski2003}, and 
$V_{\text{mic}}(\mathbf{q}, T)$ denotes the temperature-dependent 
microscopic energy corrections, calculated using the Yukawa-folded 
single-particle potential~\cite{Dobrowolski2016} and the Strutinsky 
shell correction method~\cite{Strutinsky1967, Strutinsky1968}. 
Pairing correlations are described within the BCS formalism with 
an approximate projection onto good particle number \cite{Gozdz1986,Pilat1989}. 
Further details on the macro-micro model parameters used in the present study can 
be found in Ref.~\cite{Pomorski2022}.

The temperature dependence of the microscopic corrections is described 
by the following phenomenological relation~\cite{Kostryukov2021}:
\begin{equation}
V_{\text{mic}}(\mathbf{q}, T) \approx V_{\text{mic}}(\mathbf{q}, T=0) 
\left[ 1 + \exp\left( \frac{T - 1.5}{0.3} \right) \right]^{-1},
\label{eq:vmic_temp}
\end{equation}
where the temperature $T$ is expressed in MeV. This relation ensures 
that shell and pairing corrections are gradually washed out with 
increasing excitation energy, consistently with the expected behavior 
of microscopic effects at finite temperature. 

\begin{figure}[htbp]
\centering
\includegraphics[width=0.5\textwidth,height=6.5cm]{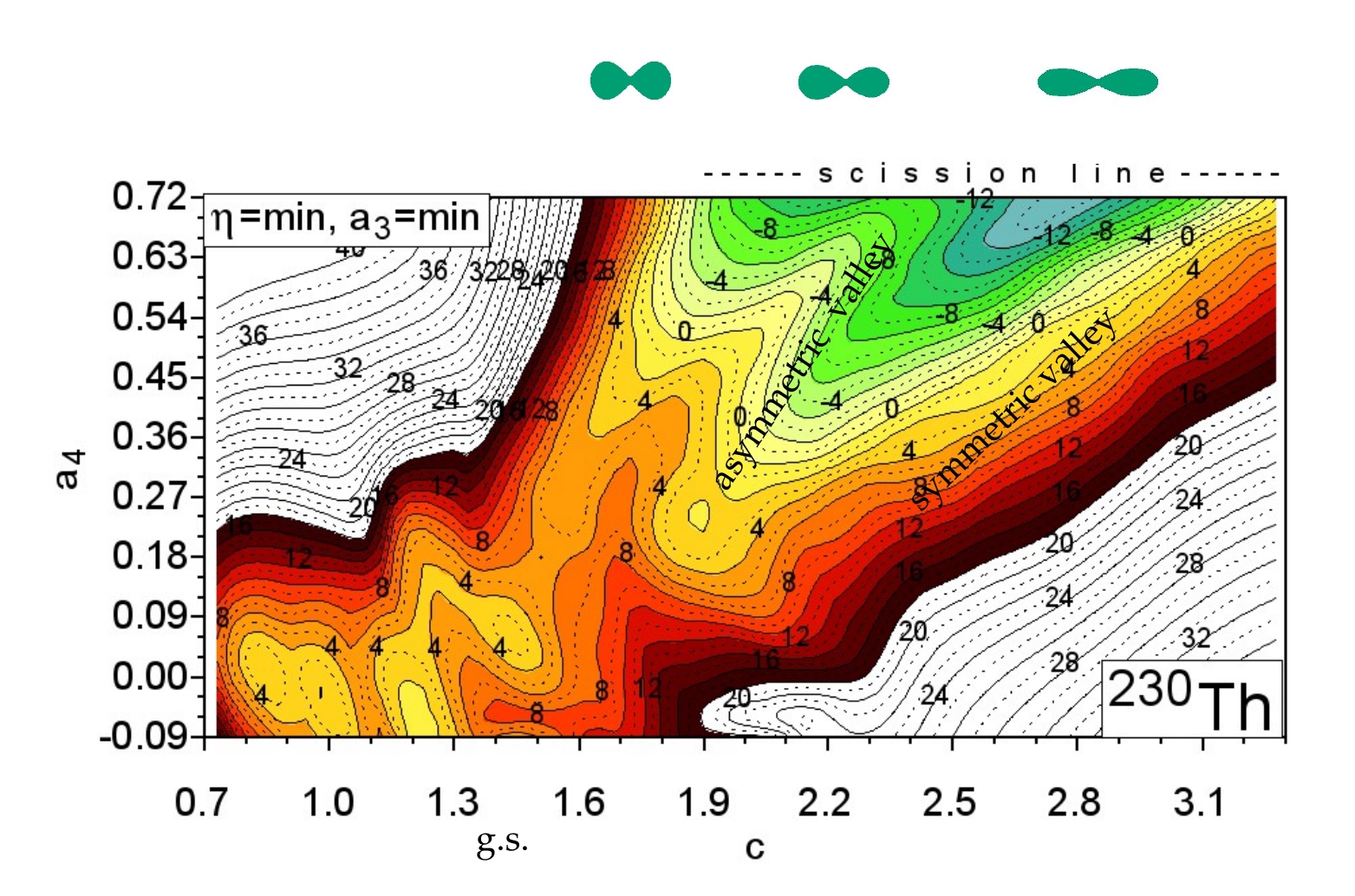} 
\caption{Potential energy surface of $^{230}$Th calculated 
within the macroscopic-microscopic model using the 4D FoS 
shape parametrization, projected onto the $(c, a_4)$ plane 
by minimization over the $a_3$ and $\eta$ coordinates. 
Contour lines represent energy levels (in MeV) relative to 
the energy of the spherical configuration. The shapes 
displayed above the surface illustrate typical nuclear 
configurations at the termination of the asymmetric and 
symmetric fission paths.}
\label{fig:pes_th230}
\end{figure}

To illustrate the energy landscape governing the fission process, 
we present in Fig.~\ref{fig:pes_th230} the PES for the nucleus $^{230}$Th at $T = 0$ MeV. 
This surface provides detailed 
information about the deformation energy and reveals the characteristic 
structures governing fission dynamics.

Figure~\ref{fig:pes_th230} shows a projection of the PES onto the plane 
defined by the elongation parameter $c$ and the neck parameter $a_4$, 
obtained by minimization over the mass-asymmetry coordinate $a_3$ and 
the non-axiality parameter $\eta$ at each point. Non-axiality 
significantly influences the energy landscape in the vicinity of the 
first saddle point, but its impact diminishes as the nucleus elongates 
towards scission---the regime most pertinent to fragment mass 
division~\cite{pomorski2024}.

The overall structure of the PES is consistent with the well-established fission energy landscape characteristic of actinide nuclei 
(see, e.g.,~\cite{Ward2017}).
One can identify the ground-state minimum at a relatively compact deformation 
(around $c \approx 1.2$). As the nucleus deforms towards fission, it 
encounters the first potential barrier (inner barrier). Beyond this 
inner barrier lies a distinct second minimum (near $c \approx 1.4$), 
representing the superdeformed fission isomer characteristic of many 
actinides. To proceed towards scission, the system must surmount the 
second potential barrier (outer saddle), located around $c \approx 1.5$ 
in this projection. The existence and properties of this double-humped 
barrier structure (the barrier heights and the depth of the isomeric 
minimum) are crucial elements determining fission probabilities and 
cross sections.

Dynamical calculations aimed at describing fragment properties are 
typically initiated near the outer saddle point, assuming the system 
has already surmounted the preceding barriers, either by quantum 
tunneling or through sufficient excitation energy, and is beginning 
its irreversible descent towards scission. The decisive choices 
determining the final fragment properties, in particular the mass 
split, are made in the region beyond this outer saddle.

Emerging from the outer saddle, the PES develops into distinct fission 
valleys, clearly visible in the $^{230}$Th example. These valleys 
represent the energetically most favorable pathways towards scission. 
Two dominant valleys can be distinguished: one corresponding to 
asymmetric fission ($a_3 \neq 0$) and one leading to symmetric fission 
($a_3 \approx 0$). For $^{230}$Th, the symmetric valley is energetically 
deeper than the asymmetric ones, indicating that symmetric fragmentation 
is the thermodynamically preferred outcome over a wide range of 
elongations beyond the saddle.

At excitation energies typical for thermal neutron capture, the system 
crosses the outer saddle with energy only slightly above the saddle 
value, marking the onset of the descent toward scission. In this regime, 
the ridge separating the fission valleys---often several MeV high--- 
effectively prevents access to the symmetric valley, making asymmetric 
fission the dominant pathway. As the excitation energy increases, the 
inter-valley ridge becomes less restrictive relative to the total 
energy, and enhanced fluctuations allow the system to sample trajectories 
leading into the symmetric valley. This provides a physically intuitive 
explanation, directly rooted in the PES structure, for the experimentally 
observed rise of the symmetric fission component with increasing 
excitation energy.

\section{Langevin Dynamics Approach}
\label{sec:langevin}

Having defined the PES on which the nuclear dynamics 
unfolds, we now turn to the two stochastic approaches used to simulate 
the fission process. We begin with the Langevin dynamics framework, 
in which the collective motion is described by coupled stochastic 
differential equations for the three deformation coordinates 
$\mathbf{q} = \{c, a_3, a_4\}$ and their conjugate momenta. 
Although the full PES is computed in the four-dimensional space 
$\{c, a_3, a_4, \eta\}$, the influence of non-axiality on the dynamics 
beyond the outer saddle point is small~\cite{pomorski2024}. Its effect 
is therefore accounted for implicitly by performing the dynamical 
calculations on a PES pre-minimized with respect to $\eta$ at each 
point in the $(c, a_3, a_4)$ space, effectively reducing the dynamical 
problem to three dimensions.

The dynamics of the system is governed by coupled Langevin equations 
for the components $q_i$ of the deformation vector $\mathbf{q}$ and 
their conjugate momenta $p_i$. The collective inertia tensor 
$m_{ij}(\mathbf{q})$, computed within the Werner-Wheeler approximation for incompressible, irrotational flow~\cite{Bartel2019}, determines the rate of change 
of the coordinates via:
\begin{equation}
\frac{d q_i}{dt} = \sum_j [m^{-1}(\mathbf{q})]_{ij} p_j.
\label{eq:langevin_q}
\end{equation}
The evolution of the momenta is governed by conservative and dissipative forces:
\begin{align}
\frac{d p_i}{dt} = & -\frac{\partial V(\mathbf{q},T)}{\partial q_i} 
- \frac{1}{2} \sum_{j,k} \frac{\partial [m^{-1}(\mathbf{q})]_{jk}}
{\partial q_i} p_j p_k \notag \\
& - \sum_{j,k} \gamma_{ij}(\mathbf{q}) [m^{-1}(\mathbf{q})]_{jk} p_k 
+ \sum_j g_{ij}(\mathbf{q}) \Gamma_j(t).
\label{eq:langevin_p}
\end{align}
The first term is the conservative force derived from the 
temperature-dependent potential energy $V(\mathbf{q}, T)$. The second 
term represents the inertia-gradient force arising from the coordinate 
dependence of the inertia tensor. The third term is the average of the dissipative force and represents 
the friction force controlled by the dissipation tensor 
$\gamma_{ij}(\mathbf{q})$, estimated within the Wall formula~\cite{blocki1978, 
Sierk1980}, whose explicit form is given later. 
The last term represents the stochastic force driven by $\Gamma_j(t)$, 
a Gaussian white noise satisfying $\langle \Gamma_i(t) \rangle = 0$ 
and $\langle \Gamma_i(t) \Gamma_j(t') \rangle = 
2\delta_{ij}\delta(t-t')$, ensuring the absence of temporal 
correlations between different time steps.

The thermodynamic temperature $T$ entering $V(\mathbf{q}, T)$ is 
determined self-consistently at each time step from the intrinsic 
excitation energy:
\begin{equation}
E_{\text{int}} = E_{\text{tot}} - \frac{1}{2} \sum_{i,j} 
[m^{-1}(\mathbf{q})]_{ij} p_i p_j - V(\mathbf{q}, T=0),
\label{eq:e_int}
\end{equation}
where the second term is the collective kinetic energy and $E_{\text{tot}}$ 
is the total energy conserved along each trajectory. The temperature 
is then obtained from the Fermi-gas relation $E_{\text{int}} = 
a(\mathbf{q}) T^2$, where $a(\mathbf{q})$ is the average single-particle 
level-density parameter~\cite{Pomorski2003}.

The strength of the random force $g_{ij}(\mathbf{q})$ in 
Eq.~(\ref{eq:langevin_p}) is related to the dissipation tensor via 
the fluctuation-dissipation theorem:
\begin{equation}
\sum_k g_{ik}(\mathbf{q}) g_{jk}(\mathbf{q}) = \gamma_{ij}(\mathbf{q}) T^*,
\label{eq:fluct_diss}
\end{equation}
where $T^*$ is a quantum-corrected effective temperature that replaces 
the classical temperature $T$ to account for zero-point fluctuations 
at low excitation energies:
\begin{equation}
T^* = \frac{E_0}{\tanh\left(E_0/T\right)},
\label{eq:tstar}
\end{equation}
with $E_0 \approx 1$~MeV being the zero-point energy of the oscillators 
composing the thermal bath~\cite{Hofmann1977, pomorski2024}. In the 
low-temperature limit ($T \to 0$), $T^* \to E_0$, so that quantum 
fluctuations set a finite lower bound on the effective temperature. 
In the high-temperature limit ($T \to \infty$), $T^* \to T$, 
recovering the classical fluctuation-dissipation theorem. 
We note that while the full quantum Langevin framework gives rise to 
colored noise with memory effects, in the present calculations we adopt 
the Markovian approximation of memoryless white noise, with quantum 
corrections incorporated solely through the effective temperature $T^*$.

The dissipation tensor $\gamma_{ij}(\mathbf{q})$ entering 
Eq.~(\ref{eq:langevin_p}) is estimated using the one-body Wall 
model~\cite{blocki1978, Sierk1980}. The underlying 
physical picture is rooted in the observation that at moderate nuclear 
excitation energies, the mean free path of a nucleon within the nucleus 
is relatively long, owing primarily to the Pauli exclusion principle, 
which suppresses two-body collisions by blocking available final states. 
Within this framework, nucleons are treated as a Knudsen gas: they move 
nearly freely inside the nuclear volume and interact with the nuclear 
medium primarily through collisions with the confining surface defined 
by the nuclear shape.

Energy dissipation arises from the momentum exchange between nucleons 
and the time-dependent moving nuclear surface. When collective motion 
causes the surface to move, the velocity of a surface element generally 
differs from the average normal velocity component of nucleons 
approaching it. Elastic reflection from this moving wall modifies the 
velocity distribution of the reflected nucleons relative to the bulk. 
The standard Wall model assumes prompt randomization of nucleon momenta 
after each wall collision, ensuring that the incident flux always 
represents the bulk distribution. Under this assumption, the friction 
tensor elements are given by~\cite{Sierk1980}:
\begin{equation}
\gamma_{ij}^{\text{wall}}(\mathbf{q}) = \frac{1}{2} \pi \rho_m \bar{v} 
\int_{z_{\text{min}}}^{z_{\text{max}}} dz\, 
\frac{\partial \rho_s^2}{\partial q_i} 
\frac{\partial \rho_s^2}{\partial q_j} 
\left[ \rho_s^2 + \frac{1}{4}
\left(\frac{\partial \rho_s^2}{\partial z}\right)^2 
\right]^{-1/2},
\label{eq:wall_friction}
\end{equation}
where $\rho_m$ is the nuclear matter density, $\bar{v} = \frac{3}{4} 
v_f$ is the average nucleon speed expressed in terms of the Fermi 
velocity $v_f$, and $\rho_s(z)$ describes the nuclear surface profile 
given by Eq.~(\ref{eq:fos_rho}) under the assumption of axial symmetry. 
The derivatives are taken with respect to the collective coordinates 
$q_i$, $q_j$ and the symmetry axis coordinate $z$.

The Wall model expression in Eq.~(\ref{eq:wall_friction}) represents 
the theoretical upper limit of the one-body dissipation 
associated with surface motion, 
corresponding to fully developed viscous flow in which all 
nucleons at the Fermi surface participate coherently in 
damping collective motion. However, 
experimental evidence indicates that the effective nuclear friction in 
neutron-induced fission remains well below this limit and exhibits a 
marked temperature dependence. At low excitation energies, pairing 
correlations suppress nucleon-nucleon collisions and the system 
operates in a nearly collisionless, ballistic regime where friction is 
minimal. As temperature increases, collisions become more frequent and 
friction grows, yet the effective friction remains significantly below 
the Wall model prediction even at elevated excitation energies.

To account for this behavior, we adopt the phenomenological 
temperature-dependent dissipation tensor similar to the one proposed in 
Ref.~\cite{Kostryukov2021}:
\begin{equation}
\gamma_{ij}(\mathbf{q}) = \frac{\left(\frac{2}{3} - \varepsilon\right) 
\gamma^{\text{wall}}_{ij}(\mathbf{q})}{1 + \exp\left[\dfrac{1.3 - T}{0.2}\right]} 
+ \varepsilon\, \gamma^{\text{wall}}_{ij}(\mathbf{q}),
\label{eq:friction_phen}
\end{equation}
where $T$ is the local nuclear temperature in MeV and $\varepsilon = 
0.05$ ensures a small but finite residual friction at $T \to 0$. 
This functional form interpolates between the ballistic limit at low 
temperatures and a value approaching $\frac{2}{3}\gamma^{\text{wall}}(\mathbf{q})$ 
at high temperatures, consistently with microscopic calculations based 
on the relaxation time approximation and linear response theory for a 
Woods--Saxon potential~\cite{Ivanyuk1996}. 
The influence of the friction strength on the fission dynamics and fragment mass distributions will be addressed in the later part of this work.

The Wall model as given by Eq.~(\ref{eq:wall_friction}) 
captures the dissipation arising from the coupling between 
collective motion and the bulk motion of nucleons confined 
by a single moving surface, and is therefore most 
appropriate for compact, mononuclear shapes. In the late 
stages of the descent to scission, however, the formation 
of a pronounced neck connecting two nascent fragments 
introduces an additional dissipation mechanism: the 
passage of nucleons through the neck window between the 
two nascent fragments, which tends to equilibrate their 
internal momentum distributions. This contribution, known 
as the \emph{window} friction~\cite{blocki1978, 
Sierk1980}, becomes 
increasingly important as the neck radius decreases. In 
contrast to heavy-ion collisions, however, the relative 
velocity of the two nascent fragments before scission 
remains small throughout the descent from saddle to 
scission, so that the window contribution is expected to 
play a less prominent role than in the case of colliding 
nuclei. In the present work, window friction is therefore 
not included explicitly; its effect is assumed to be 
subdominant and partially absorbed into the phenomenological 
temperature dependence of Eq.~(\ref{eq:friction_phen}). A 
more complete treatment including window friction 
explicitly, and its role in shaping the late-stage 
dynamics, is deferred to future work.

The formalism defined above governs the temporal evolution of the 
system from the initial configuration to scission. 
Scission is defined by a critical neck radius approximately 
equal to the nucleon size, for which we adopt the value of
$1.2$~fm. This condition corresponds to 
$a_4 \approx 0.72$ in the FoS 
parametrization~\cite{pomorski2024}.
Trajectories are initiated at a total energy $E_{\text{tot}} = 
V(\mathbf{q}_0, T=0) + E^*$, where $E^*$ is the initial excitation 
energy measured relative to the ground state. For each fissioning system, 
an ensemble of $10^5$ independent Langevin trajectories is generated.

The choice of the starting point $\mathbf{q}_0$ warrants discussion. 
For low excitation energies typical of neutron capture reactions, 
dynamical calculations are commonly initiated in the vicinity of the 
outer saddle point or the second minimum. This is partly motivated by 
the topology of the PES, which strongly constrains low-energy 
trajectories in the region between the ground state and the outer 
saddle. The primary practical motivation, however, is computational: 
at low excitation energies, the system may spend a considerable amount 
of time in the first and second minima before surmounting the outer 
saddle, making full trajectories from the ground state prohibitively 
expensive. In the present work, all calculations are therefore 
initiated in the vicinity of the outer saddle point, whose location 
is determined using the immersion method following Ref.~\cite{AugustynPRC}. 
The only exception is fission of $^{256}$Fm, for which two competing shallow outer 
saddles are present on the PES. In this case, the calculations are 
initiated from the second minimum, allowing the system to naturally 
select between the available fission pathways.

\begin{figure}[htbp] 
\centering
\includegraphics[width=0.5\textwidth]{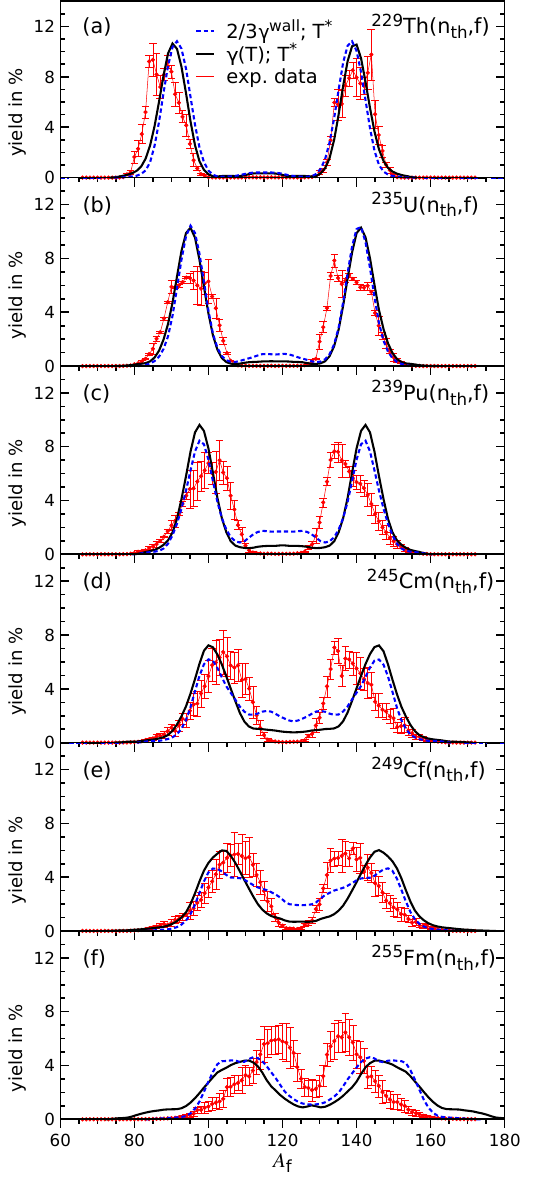} 
\caption{Pre-neutron emission fragment mass distributions for 
thermal neutron-induced fission of selected actinide 
nuclei from $^{229}$Th to $^{255}$Fm, calculated using the 
Langevin model. Solid black curves show results obtained with the 
phenomenological temperature-dependent dissipation tensor of 
Eq.~(\ref{eq:friction_phen}), while dashed blue curves correspond 
to its upper limit, 
$\gamma_{ij}(\mathbf{q}) = \tfrac{2}{3}\,
\gamma^{\mathrm{wall}}_{ij}(\mathbf{q})$. In both cases, 
the quantum-corrected effective temperature $T^*$ of 
Eq.~(\ref{eq:tstar}) is used in the fluctuation-dissipation 
relation. Experimental post-neutron emission data (symbols) 
are taken from ENDF/B-VIII.0~\cite{NNDC}.}
\label{fig:langevin_systematics}
\end{figure}

To illustrate the capabilities and systematic trends predicted by the 
Langevin dynamics model, Fig.~\ref{fig:langevin_systematics} presents 
a comparison of calculated pre-neutron-emission fragment mass distributions 
with experimental data for a series of thermal neutron-induced fission 
reactions on actinide targets spanning from $^{229}$Th to $^{255}$Fm. 
All experimental data are taken from the evaluated nuclear data library 
ENDF/B-VIII.0~\cite{NNDC}. The excitation energies involved in the 
considered processes are sufficiently low that the additional complexity 
of multi-chance fission contributions is avoided, allowing for a clean 
assessment of the model's predictive performance across different nuclear 
systems within a consistent and well-controlled theoretical framework. 
Additionally, neither pre-scission nor post-scission neutron emission is 
included in the present treatment. This simplification is justified by 
the simplicity of the model and, in the case of pre-scission emission, by 
the fact that, even where energetically allowed, such emission could only 
occur during the very short time interval preceding scission.

The calculated pre-neutron emission mass distributions are 
shown as black lines in Fig.~\ref{fig:langevin_systematics}, 
while the experimental data correspond to post-neutron 
emission yields. Pre-neutron distributions provide a more 
direct view of the primary model output, facilitating 
systematic comparisons across the actinide chain. It is 
worth noting that neutron evaporation does not significantly 
alter the leading peak positions or relative 
heights~\cite{pomorski2024}, indicating that 
the primary features of the mass distribution are 
established before significant fragment de-excitation.

For thorium, the model reproduces both the centroids and widths 
of the fragment mass peaks with the best overall agreement. Moving 
towards heavier systems, the experimental distributions exhibit a 
systematic broadening and convergence of the two peaks, reflecting the 
well-known trend towards more symmetric mass splits. The model captures 
this qualitative trend, but with quantitative discrepancies: the 
predicted peak centroids are more widely separated than observed 
experimentally, with the heavy fragment peak systematically 
overestimated and the light fragment peak underestimated.

At the low excitation energies characteristic of thermal neutron 
capture, the phenomenological dissipation tensor of 
Eq.~(\ref{eq:friction_phen}) yields values well below the Wall model 
limit in the region between the ground state and the outer saddle, 
consistent with the ballistic regime expected at low nuclear 
temperatures. Beyond the outer saddle, as the potential energy 
decreases and the available thermal energy increases, the dissipation 
grows progressively and becomes most pronounced in the vicinity of 
scission.
Even so, the Langevin model predicts a small but non-zero 
symmetric fission component. This is a direct 
consequence of the stochastic nature of the Langevin equations: the 
random force term, incorporating both thermal statistical and quantum 
zero-point fluctuations via the effective temperature $T^*$, provides 
a mechanism for barrier crossing, allowing a fraction of trajectories 
to populate the symmetric fission valley even when the average energy 
of the system appears insufficient to surmount the inter-valley ridge 
on the PES. This tendency is further enhanced across the actinide 
chain: as one moves from thorium towards heavier systems such as fermium, the 
ridge separating the asymmetric and symmetric fission valleys on the 
PES progressively diminishes, making the transition to the symmetric 
valley increasingly accessible. This finite symmetric yield, absent 
in purely deterministic models, reflects an essential physical feature 
of the stochastic approach. 

To further illustrate the role of friction, Fig.~\ref{fig:langevin_systematics} 
also shows results obtained with the upper limit of Eq.~(\ref{eq:friction_phen}), 
$\gamma_{ij}(\mathbf{q}) = \frac{2}{3}\gamma_{ij}^{\text{wall}}(\mathbf{q})$ (blue solid 
lines). In this high-friction regime, the strong damping is accompanied 
by correspondingly large random forces, leading to enhanced diffusion 
across the PES. As a result, the symmetric fission component is 
overestimated relative to experiment, with trajectories 
readily crossing the inter-valley ridge towards symmetric 
configurations. Conversely, with the phenomenological friction of 
Eq.~(\ref{eq:friction_phen}), the reduced damping allows the 
deterministic potential gradient to play a more dominant role: once 
the system begins descending into an asymmetric valley, stochastic 
deflection towards the symmetric valley becomes less likely. 
This behavior is consistent with the picture of fission dynamics in the 
descent from saddle to scission operating mostly in a moderately damped 
regime, where the topology of the PES governs the outcome more strongly 
than diffusive effects. The results obtained with the phenomenological friction are in considerably 
better agreement with experimental data, supporting this interpretation, 
with the exception of fermium, where a significant fraction of trajectories 
reaches more asymmetric configurations. 
However, it should be noted that the present Langevin framework incorporates 
both thermal and quantum fluctuations, and their interplay may play a 
non-negligible role in shaping the final mass distributions. The limit of 
strong friction without additional quantum fluctuations will be discussed 
in more detail in the following section, allowing for a cleaner assessment 
of the role of diffusive effects in the overdamped regime.

\section{Metropolis Random Walk Approach}
\label{sec:randomwalk}
The random walk approach can be formally derived from the Langevin 
equations in the overdamped limit of strong friction~\cite{Randrup2011_PRC84}, in which 
the inertial term in Eq.~(\ref{eq:langevin_p}) becomes negligible 
and the momenta cease to be independent dynamical variables. The 
evolution then reduces to a diffusive process in coordinate space 
governed by the potential gradient and stochastic forces, as 
described by the Smoluchowski equation. In this regime, the system 
has no memory of its previous momenta and the transition probabilities 
depend only on the current position $\mathbf{q}$ on the PES, which 
is precisely the assumption underlying the random walk model.

To investigate this limit, we employ the Metropolis-type walk methodology~\cite{10.1063/1.1699114} based 
on the works of~\cite{randrup2011, Randrup2011_PRC84, Randrup2013}, applied here as a complementary stochastic 
approach that explicitly probes the overdamped regime. The calculations 
are performed within the same three-dimensional FoS deformation space 
$\{c, a_3, a_4\}$ and on the same $\eta$-minimized PES described in 
Sec.~\ref{sec:framework}, ensuring a consistent basis for comparison 
with the Langevin dynamics results.

In the adopted random walk approach, the state of the system is therefore 
represented solely by its position $\mathbf{q}$ on a discretized 
deformation grid, and the evolution proceeds as a sequence of 
probabilistic steps between neighboring grid points. The transition 
probability is governed by the density of intrinsic nuclear states 
$\rho(E_{\text{int}}(\mathbf{q}))$ at each configuration, where the intrinsic 
excitation energy is defined as:
\begin{equation}
E_{\text{int}}(\mathbf{q}) = E_{\text{tot}} - V(\mathbf{q}, T),
\label{eq:eint_rw}
\end{equation}
without a collective kinetic energy term, consistently with the 
absence of explicit momenta in the model. The random walk thus probes to what extent the fragment
mass distribution is determined by the entropy landscape
$S(\mathbf{q}) = \ln \rho(E_{\mathrm{int}}(\mathbf{q}))$ associated with the
PES, in the absence of inertial and dissipative effects.

In Eq.~(\ref{eq:eint_rw}), the potential energy $V(\mathbf{q}, T)$ 
is computed in the same way as in the Langevin calculations 
(see Eqs.~(\ref{eq:pes_total}) and~(\ref{eq:vmic_temp})). 
Since $V(\mathbf{q}, T)$ depends 
explicitly on the nuclear temperature $T$, which is itself determined 
by $E_{\text{int}}(\mathbf{q})$, the evaluation of the available energy 
proceeds iteratively until self-consistency is achieved. 
This differs from the approach of Ref.~\cite{Randrup2013}, 
where shell effects are instead incorporated into the level 
density via the Ignatyuk prescription~\cite{Ignatyuk}. The 
two formulations are physically equivalent but differ in 
the functional form and the characteristic scale of the 
shell damping; at the low excitation energies relevant 
here, this difference is not expected to significantly 
affect the transition probabilities. Consequently, the 
level density adopted here can be approximated by the 
simple Fermi gas model without additional damping terms:
\begin{equation}
\rho(E_{\text{int}}(\mathbf{q})) \propto 
\exp\left(2\sqrt{a_0\, E_{\text{int}}(\mathbf{q})}\right),
\label{eq:dos_rw}
\end{equation}
where the pre-exponential factor is neglected when computing relative 
transition probabilities. The present calculations employ a constant level density parameter 
$a_0 = A/8.5$~MeV$^{-1}$, a standard value widely used in 
macroscopic-microscopic approaches. While this approximation is well 
established, the level density parameter is in general 
deformation- and energy-dependent~\cite{Rahmatinejad2021, 
Rahmatinejad2022}, and recent studies have shown that shell effects 
and collective modes can significantly alter the available level 
density along the fission path~\cite{Rahmatinejad2024}. For the 
relatively low excitation energies considered in the present work, 
however, these effects are expected to be moderate, and the constant 
parameter approximation remains reasonable. This is further supported 
by combinatorial level density calculations, which show no significant 
deviation from the Fermi gas model in this energy regime~\cite{Ward2017}.

Following the Metropolis-type prescription of 
Ref.~\cite{Randrup2013}, the evolution proceeds as follows. 
At each step, a candidate neighboring point $\mathbf{q}_j$ is 
first selected at random, with uniform probability, from the 
set of $26$ nearest neighbors of the current point 
$\mathbf{q}_i$ on the three-dimensional cubic grid. These correspond to all combinations of displacements 
$(\delta c, \delta a_3, \delta a_4)$ with 
$\delta c \in \{-\Delta c, 0, +\Delta c\}$, 
$\delta a_3 \in \{-\Delta a_3, 0, +\Delta a_3\}$, and 
$\delta a_4 \in \{-\Delta a_4, 0, +\Delta a_4\}$, excluding 
the null displacement. Here, $\Delta c$, $\Delta a_3$, 
and $\Delta a_4$ denote the 
grid spacings along the respective coordinates.
This Moore-type neighborhood, in 
contrast to the more restrictive axial (von Neumann) 
neighborhood of six sites, permits diagonal moves in the 
deformation space and thereby ensures a more efficient 
sampling, particularly across the ridge separating the 
symmetric and asymmetric fission valleys. A candidate is 
considered accessible only if 
$E_{\mathrm{int}}(\mathbf{q}_j) > 0$, i.e., if the total 
energy is sufficient to reach that configuration; otherwise 
it is rejected and a new candidate is drawn. The accessible 
candidate is then accepted or rejected according to the 
Metropolis criterion based on the ratio of level densities 
at the two sites:
\begin{equation}
P_{i \to j} = \min\!\left[\, 1,\, 
\frac{\rho(E_{\mathrm{int}}(\mathbf{q}_j))}
{\rho(E_{\mathrm{int}}(\mathbf{q}_i))}\right].
\label{eq:rw_prob}
\end{equation}
The unity appearing in Eq.~(\ref{eq:rw_prob}) reflects the 
fact that any move towards a configuration of higher level 
density, $\rho(E_{\mathrm{int}}(\mathbf{q}_j)) > 
\rho(E_{\mathrm{int}}(\mathbf{q}_i))$, is accepted with 
certainty, whereas moves towards lower level density are 
accepted only with the probability given by the ratio. 
This asymmetry drives the system towards regions of high 
density of intrinsic states --- equivalently, towards lower 
potential energy --- while still permitting fluctuations 
against the gradient, which are essential for ergodic 
sampling and for the crossing of inter-valley ridges on 
the PES. If the candidate is rejected, the current site 
$\mathbf{q}_i$ is retained and counted as the next step in 
the sequence. By construction, this prescription satisfies 
detailed balance with respect to the weight 
$\rho(E_{\mathrm{int}}(\mathbf{q}))$, guaranteeing that the 
stationary distribution sampled by the walk is proportional 
to the density of intrinsic states and is thus consistent 
with the microcanonical ensemble underlying the model.

This choice represents a methodological refinement over the 
proportional sampling scheme applied in our earlier study of 
fusion dynamics~\cite{Cap2024}, in which the transition 
probability was taken directly proportional to the level 
density at the destination, 
$P_{i \to j} \propto \rho(E_{\mathrm{int}}(\mathbf{q}_j))$. 
Beyond the formal advantage of satisfying detailed balance, 
the two prescriptions differ in how they explore the 
deformation space. Proportional sampling weights each 
candidate by the magnitude of the local increase in $\rho$, 
thereby channeling trajectories along the steepest ascent of 
the level density --- equivalently, along the steepest 
descent of the potential energy. The Metropolis prescription, 
by contrast, accepts any move towards higher $\rho$ with 
unit probability regardless of the magnitude of the increase, 
which leads to a broader exploration of the deformation space 
around the dominant PES gradient and, consequently, to a 
more realistic width of the resulting fragment mass 
distributions.

An important practical aspect of the random walk implementation 
is the choice of grid spacing $\Delta q_i$, particularly for the 
mass-asymmetry coordinate $a_3$, since the resolution of the 
predicted mass distribution is directly determined by 
$\Delta a_3$. The fragment masses at scission are determined 
geometrically from the nuclear shape at the termination point 
of each trajectory ($a_4 \approx 0.72$), by calculating the 
mass on each side of the scission neck. For the purpose of 
estimating the required grid resolution, one may use the 
approximate linear relation 
$A_h \approx \frac{A}{2}(1 + 0.9894\, a_3)$, valid near 
scission~\cite{pomorski2024}, where $A_h$ denotes the mass 
number of the heavy fragment and $A$ is the mass number of 
the fissioning nucleus. This relation shows that a step 
$\Delta a_3 = 0.01$ corresponds to approximately one mass 
unit, which sets the natural scale for the grid resolution.

The choice $\Delta a_3 = 0.01$ is dictated primarily by this 
mass-resolution requirement: a larger step would degrade the 
accuracy of the predicted mass distribution, while a 
significantly smaller step offers no practical gain. The 
remaining coordinates $c$ and $a_4$ are less sensitive in 
this respect, as they are not directly tied to the mass 
split. A typical variation of $\Delta a_3 = 0.01$ corresponds 
to a change in the potential energy of approximately $1$~MeV; 
the same magnitude of energy variation is obtained for steps 
of $0.01$ in $c$ and $a_4$, which led us to adopt a uniform 
grid spacing $\Delta c = \Delta a_3 = \Delta a_4 = 0.01$. 
Test calculations performed with alternative step sizes did 
not reveal any significant impact on the resulting fragment 
mass distributions, in line with the observations of 
Ref.~\cite{Randrup2013}.

As in the Langevin approach, trajectories are initiated at the outer 
saddle point, or at the second minimum in the case of fermium, 
with total energy $E_{\mathrm{tot}} = V(\mathbf{q}_0, T=0) + E^*$, 
and terminated upon reaching the scission condition.
For each fissioning system, an ensemble of $10^5$ 
independent random walks is generated, providing 
statistics consistent with the Langevin calculations.
To obtain a continuous distribution of mass splits despite 
the discreteness of the underlying grid, the final deformation 
parameters reached at scission are randomized within a uniform 
sub-grid window: each component $q_i$ is replaced by 
$q_i + \delta_i$, with $\delta_i$ drawn independently from a 
uniform distribution on 
$[-\Delta q_i / 2, +\Delta q_i / 2]$. The fragment masses 
are then computed geometrically from 
the randomized shape, in the same way as in the Langevin 
approach, and rounded to the nearest integer value. This 
procedure removes the spurious discretization of the mass 
yield inherited from the lattice and yields smooth, 
physically meaningful mass distributions without altering 
the underlying stochastic dynamics.

\begin{figure}[htbp] 
\centering
\includegraphics[width=\columnwidth]{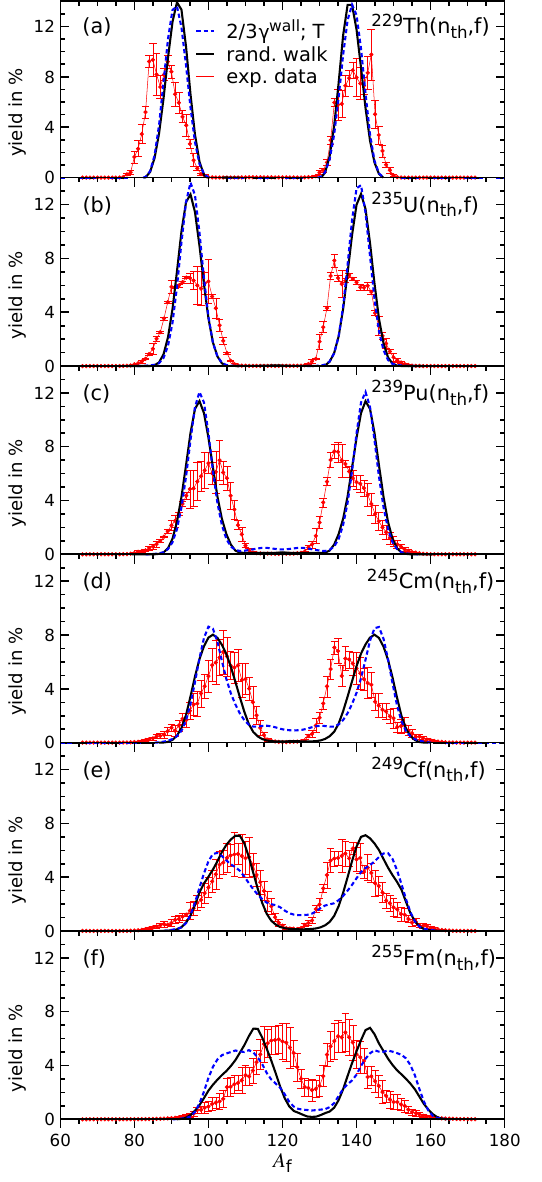} 
\caption{Pre-neutron emission fragment mass distributions for 
thermal neutron-induced fission of selected actinide 
nuclei from $^{229}$Th to $^{255}$Fm, calculated using the 
Metropolis random walk model (solid black curves). For comparison, 
pre-neutron emission Langevin calculations performed in 
the high-friction limit 
($\gamma_{ij}(\mathbf{q}) = \tfrac{2}{3}\,
\gamma^{\mathrm{wall}}_{ij}(\mathbf{q})$), with the 
classical temperature $T$ replacing the effective 
temperature $T^*$ in Eq.~(\ref{eq:fluct_diss}), are also 
shown (dashed blue curves). Experimental post-neutron emission 
data (symbols) are taken from ENDF/B-VIII.0~\cite{NNDC}.}
\label{fig:random_walk_systematics}
\end{figure}

Figure~\ref{fig:random_walk_systematics} presents the primary 
(pre-neutron emission) fragment mass distributions calculated 
with the Metropolis random walk model (black lines) for thermal 
neutron-induced fission of selected actinide nuclei. The calculated distributions are 
compared with experimental post-neutron emission data taken 
from ENDF/B-VIII.0~\cite{NNDC} (symbols). As discussed in 
Sec.~\ref{sec:langevin}, neutron evaporation does not 
significantly alter the leading peak positions or relative 
heights, so that pre-neutron distributions can be meaningfully 
compared with post-neutron data without introducing 
substantial bias. 

For reference, Langevin calculations (blue lines) performed 
in the high-friction limit 
$\gamma_{ij}(\mathbf{q}) = \tfrac{2}{3}\,
\gamma^{\mathrm{wall}}_{ij}(\mathbf{q})$ are also shown, 
allowing a direct assessment of the correspondence between 
the two approaches in the overdamped regime. The choice 
$\tfrac{2}{3}\gamma^{\mathrm{wall}}$ is motivated physically: 
it corresponds to the high-temperature limit of the 
phenomenological dissipation tensor of 
Eq.~(\ref{eq:friction_phen}) and is consistent with 
microscopic calculations of one-body dissipation based on 
linear response theory~\cite{Ivanyuk1996}. It 
therefore represents a physically motivated upper bound on 
the effective nuclear friction rather than an abstract 
$\gamma \to \infty$ limit. While this damping is strong 
enough to suppress the inertial term in 
Eq.~(\ref{eq:langevin_p}) to a large extent, it is 
milder than the strict overdamped regime in which the 
momenta are fully equilibrated on timescales much shorter 
than the coordinate evolution. At the low excitation 
energies characteristic of thermal neutron capture this 
distinction is not expected to be decisive, but it should 
be kept in mind when the formalism is applied to fission 
from more highly excited states, where the residual 
inertial dynamics may play a more significant role.
In these reference calculations, the classical temperature $T$ 
replaces the effective temperature $T^*$ in 
Eq.~(\ref{eq:fluct_diss}), so that quantum zero-point 
fluctuations are switched off. This choice ensures a 
consistent comparison with the random walk model, which, 
being derived from the classical Smoluchowski equation as 
the overdamped limit of the classical Langevin dynamics, 
samples the Fermi-gas level density 
$\rho(E_{\mathrm{int}})$ without any quantum correction. 
The correspondence probed here is therefore between two 
genuinely classical, overdamped descriptions of the 
fission dynamics.

Both the random walk and the overdamped Langevin calculations 
qualitatively reproduce the general shape of the fragment mass 
distributions observed across the actinide chain, including the 
dominant asymmetric character of the yields and the systematic 
convergence of the heavy and light peaks when moving from thorium 
towards fermium. Quantitatively, however, two systematic 
discrepancies are visible. First, the calculated peaks are 
noticeably narrower than the experimental ones for all studied 
nuclei. We attribute this narrowness primarily to the restriction 
of the dynamics to a three-dimensional deformation space 
$\{c, a_3, a_4\}$, which does not capture all the relevant 
collective modes contributing to the width of the mass 
distribution. In this respect, it is instructive to note that 
the 5D Metropolis walk calculations of 
Refs.~\cite{randrup2011, Randrup2011_PRC84, Randrup2013}, performed in a richer deformation space, 
yield broader distributions in closer agreement with experiment. 
Second, the separation between the light- and heavy-fragment 
peaks is somewhat underestimated for thorium and 
progressively overestimated towards the heavier systems. This 
systematic shift is rooted in the structure of the 
underlying PES: the position of the scission ridge in the 
deformation space determines the final mass split, and 
augmenting the parametrization with additional shape degrees 
of freedom is expected to modify this ridge and thereby the 
predicted peak positions.

The comparison between the random walk and the overdamped Langevin 
results reveals an excellent mutual agreement for the lighter 
actinides: for both $^{229}$Th and $^{235}$U, the two 
approaches produce essentially identical mass distributions, with 
coincident peak positions and widths. A small but visible deviation 
appears for $^{239}$Pu, and the discrepancy grows further for 
$^{245}$Cm, $^{249}$Cf, and $^{255}$Fm where the Langevin calculations yield a 
non-negligible symmetric fission component that is absent in the 
random walk predictions. This pattern can be understood from the 
evolution of the PES topology along the actinide chain. For the 
lighter systems, the ridge separating the asymmetric and symmetric 
fission valleys beyond the outer saddle is sufficiently pronounced 
that neither the accumulated stochastic kicks of the overdamped 
Langevin dynamics nor the Metropolis acceptance of uphill moves 
can efficiently drive trajectories across it; in both models, the 
descent follows the steepest gradient of the PES, and the two 
approaches therefore converge. As $Z$ increases towards fermium, 
this inter-valley ridge gradually diminishes, and the strong random 
forces characteristic of the high-friction Langevin regime --- 
whose magnitude scales as $\sqrt{\gamma T}$ --- become capable of 
pushing a fraction of trajectories across the ridge into the 
symmetric valley. The Metropolis walk, in contrast, explores the 
deformation space one nearest-neighbor step at a time, so that the 
probability of traversing a multi-step uphill segment is the 
product of individual small acceptance probabilities and remains 
strongly suppressed. As noted above, the adopted friction strength 
$\tfrac{2}{3}\,\gamma^{\mathrm{wall}}_{ij}(\mathbf{q})$ 
corresponds to a strongly damped but not strictly 
overdamped regime; the residual inertial dynamics retained 
in the Langevin framework, and absent in the random walk 
by construction, may therefore additionally contribute to 
the differences observed for the heavier actinides.
The absence of the symmetric component in the 
random walk predictions, consistent with experimental 
observations at these excitation energies, thus reflects the 
combined effect of the stricter kinematic constraints of the 
Metropolis sampling and the purely configurational nature of 
the random walk evolution.

\section{Role of Quantum Zero-Point Fluctuations}
\label{sec:zpf}

The parallel Langevin calculations presented in 
Figs.~\ref{fig:langevin_systematics} 
and~\ref{fig:random_walk_systematics}, performed with the 
same dissipation tensor 
$\gamma_{ij}(\mathbf{q}) = \tfrac{2}{3}\,
\gamma^{\mathrm{wall}}_{ij}(\mathbf{q})$ but with different 
prescriptions for the noise strength, allow a direct 
assessment of the contribution of quantum zero-point 
fluctuations to the fragment mass distribution. In 
Fig.~\ref{fig:langevin_systematics}, the quantum-corrected 
effective temperature $T^*$ of Eq.~(\ref{eq:tstar}) is 
used in the fluctuation-dissipation relation, 
Eq.~(\ref{eq:fluct_diss}); in 
Fig.~\ref{fig:random_walk_systematics}, it is replaced by 
the classical temperature $T$. Since all other ingredients 
of the calculation are identical, the difference between 
the two sets of results isolates the effect of the 
zero-point contribution encoded in $T^*$.

The comparison reveals that the inclusion of zero-point 
fluctuations through $T^*$ substantially enhances the 
population of the symmetric fission valley. The effect is 
already visible for $^{229}$Th and $^{235}$U (see Fig.~\ref{fig:langevin_systematics}) 
and grows rapidly 
along the actinide chain, becoming dominant for 
$^{245}$Cm and $^{249}$Cf, where the high-friction 
Langevin calculations with $T^*$ yield a symmetric 
component that clearly exceeds the experimental yields. 
Replacing $T^*$ by $T$ substantially suppresses this 
component and brings the calculated distributions into 
closer agreement with the random walk predictions and, 
qualitatively, with the experimental trend of suppressed 
symmetric fission at thermal excitation energies. 
Furthermore, the inclusion of the quantum-corrected effective temperature 
$T^*$ leads to a significant broadening of the mass distribution for 
$^{255}$Fm.

Taken together, these observations suggest that the 
standard prescription $T^* = E_0/\tanh(E_0/T)$, while 
correct in the $T \to 0$ limit where $T^* \to E_0$ as 
required by the quantization of collective oscillations, 
may sustain the zero-point contribution at intermediate 
temperatures more strongly than is physically warranted 
for fission dynamics. In other words, the physical 
damping of quantum fluctuations with increasing excitation 
energy may proceed more rapidly than the smooth 
interpolation provided by the hyperbolic tangent 
formulation. In practice, the phenomenological 
temperature-dependent friction of 
Eq.~(\ref{eq:friction_phen}) offers an effective 
compensation for this behavior: by reducing the magnitude 
of the dissipative force at low 
temperatures, it limits the impact of the $T^*$ 
correction in the regime where zero-point effects would 
otherwise be overestimated. 

A dedicated investigation of the temperature dependence 
of zero-point fluctuations in the collective fission 
dynamics --- in particular, of the interplay between the 
effective temperature $T^*$ and the strength and 
temperature profile of the dissipation tensor --- would be 
a natural extension of the present work.

\section{Final Remarks}
The results presented in this work demonstrate that two 
stochastic approaches to induced fission --- 
multidimensional Langevin dynamics in the high-friction 
regime and the Metropolis random walk model --- implemented 
within an identical 4D FoS-parameterized PES framework, 
yield consistent predictions for fragment mass distributions 
in thermal neutron-induced fission of $^{229,232}$Th, 
$^{235}$U, $^{239}$Pu, $^{245}$Cm, $^{249}$Cf, and $^{255}$Fm.
Both approaches qualitatively reproduce the dominant 
experimental trends: the asymmetric character of the mass 
split, the positions of the light- and heavy-fragment 
peaks, their systematic evolution along the chain, and the 
suppression of symmetric fission at the low excitation 
energies characteristic of thermal neutron capture.

The observed convergence of the two models for the lighter 
actinides, and its gradual breakdown towards $^{245}$Cm, 
$^{249}$Cf, and $^{255}$Fm is consistent with theoretical expectations for 
the overdamped regime. In both approaches, the descent from 
the outer saddle to scission is governed primarily by the 
topology of the PES and the associated density of intrinsic 
states, with the stochastic component acting as a 
perturbation that allows the system to explore 
configurations away from the steepest-gradient path. The 
residual differences between the two frameworks reflect the 
distinct roles of inertia and dissipation. In the Langevin 
description, the collective inertia tensor 
$m_{ij}(\mathbf{q})$, together with its coordinate 
dependence and off-diagonal couplings between deformation 
modes, imparts a finite memory to individual trajectories 
even at the adopted high-friction strength 
$\tfrac{2}{3}\gamma^{\mathrm{wall}}_{ij}(\mathbf{q})$; this 
inertial content is absent in the Metropolis walk by 
construction. Similarly, the adopted dissipation tensor, designed to 
capture the dominant one-body dissipation through the Wall 
model and its phenomenological temperature dependence, 
does not include the window contribution explicitly. Its 
effect, expected to be modest in the regime considered 
here, is absorbed implicitly into the phenomenological 
parametrization. A more complete treatment including 
window friction explicitly would allow a finer assessment 
of the role of dissipation near scission and is a natural 
extension of the present framework.

A persistent quantitative limitation common to both 
approaches concerns the widths of the mass peaks. The 
calculated distributions are systematically narrower than 
the experimental ones, indicating that the 3D dynamics 
employed here, although sufficient to capture the dominant 
features of the descent from saddle to scission and the 
systematic trends along the actinide chain, does not 
fully account for the collective modes that contribute to 
the width of the mass distribution in a more comprehensive 
parametrization. Exploratory calculations in 
higher-dimensional deformation spaces, including 
additional multipole parameters such as $a_5$ and $a_6$, 
would clarify to what extent the remaining quantitative 
discrepancies reflect genuine physical effects or 
artifacts of the restricted parametrization. A further 
natural refinement concerns the temperature dependence of 
the microscopic corrections and of the dissipation tensor. 
The phenomenological factor of Eq.~(\ref{eq:vmic_temp}) 
represents an empirical average over the distinct damping 
behavior of proton pairing, neutron pairing, and shell 
corrections, each of which is expected to fade with 
excitation energy on its own characteristic scale; a 
dedicated microscopic analysis of these individual 
scaling laws is presented in a companion 
work~\cite{Pomorski2026}. Replacing the single 
phenomenological suppression factor by component-specific 
functions, and coupling the temperature dependence of the 
dissipation tensor of Eq.~(\ref{eq:friction_phen}) to that 
of the pairing correlations --- which are known to 
suppress nucleon-nucleon collisions and thereby reduce 
the magnitude of one-body dissipation at low temperatures 
--- would eliminate the adjustable parameters currently 
entering Eqs.~(\ref{eq:vmic_temp}) 
and~(\ref{eq:friction_phen}), yielding a more predictive 
description of the interplay between shell damping, 
pairing, and dissipation in the descent to scission.

These refinements are particularly relevant in view of 
the widespread use of random walk and Brownian shape 
motion models in the description of fragment mass 
distributions for heavy~\cite{mumpower2020, MollerRandrup2015_Brownian} and superheavy 
nuclei~\cite{Albertsson2020EPJA}, 
where systematic comparisons with full Langevin dynamics 
within a common theoretical framework would help delineate 
the regime of validity of the random walk approximation 
and identify the nuclear systems for which inertial and 
dissipative effects play a decisive role. In this 
context, the connection established here between the 
Langevin and random walk frameworks also opens the way to 
hybrid modeling strategies combining the computational 
efficiency of the Metropolis approach with the full 
dynamical content of the Langevin equations --- for 
instance, by using the random walk to pre-sample the 
relevant regions of configuration space before launching 
fully dynamical trajectories. Such approaches may prove particularly useful for 
systematic surveys across the nuclear chart, where the 
efficiency and simplicity of the Metropolis sampling can 
be combined with the full dynamical content of the 
Langevin description.

\bibliographystyle{apsrev4-1} 
\bibliography{references} 

\end{document}